\documentclass[preprint]{aastex}
\usepackage{psfig}

\shortauthors{J.J.M. in 't Zand et al.}
\shorttitle{X-ray afterglow detection of GRB~991014}

\begin{document}

\slugcomment{Accepted for Astrophysical Journal part I, 2000 July 18}

\title{X-RAY AFTERGLOW DETECTION OF THE SHORT GAMMA-RAY BURST 991014}
\author{J.J.M.~in~'t~Zand\altaffilmark{1,2}, 
 L.~Kuiper\altaffilmark{1},
 L.~Amati\altaffilmark{3},
 L.A.~Antonelli\altaffilmark{4,5}
 K.~Hurley\altaffilmark{6},
 A.~Coletta\altaffilmark{7},
 E.~Costa\altaffilmark{8},
 M.~Feroci\altaffilmark{8},
 F.~Frontera\altaffilmark{3,9},
 G.~Gandolfi\altaffilmark{8},
 J.~Heise\altaffilmark{1},
 E.~Kuulkers\altaffilmark{1,10},
 J.M.~Muller\altaffilmark{1,5},
 L.~Nicastro\altaffilmark{11},
 L.~Piro\altaffilmark{7},
 M.J.S.~Smith\altaffilmark{1,7},
 M.~Tavani\altaffilmark{12,13}}

\affil{}
\altaffiltext{1}{Space Research Organization Netherlands, 
	Sorbonnelaan 2, 3584 CA Utrecht, the Netherlands}
\altaffiltext{2}{email jeanz@sron.nl}
\altaffiltext{3}{Istituto di Tecnologie e Studio delle Radiazioni Extraterrestri (CNR), Via Gobetti 101, 40129 Bologna, Italy}
\altaffiltext{4}{Osservatorio Astronomico di Roma, Via Frascati 33, 00040 Monteporzio Catone, Italy}
\altaffiltext{5}{BeppoSAX Science Data Center, Via Corcolle 19, 00131 Rome, Italy}
\altaffiltext{6}{University of California, Space Sciences Laboratory, Berkeley, CA 94720-7450, U.S.A.}
\altaffiltext{7}{BeppoSAX Scientific Operation Center, Via Corcolle 19, 00131 Rome, Italy}
\altaffiltext{8}{Istituto di Astrofisica Spaziale (CNR), 00133 Rome, Italy}
\altaffiltext{9}{Dipartimento Fisica, Universit\'{a} di Ferrara, Via Paradiso 12, 44100 Ferrara, Italy}
\altaffiltext{10}{Astronomical Institute, Utrecht University, P.O. Box 80 000, 3504 TA Utrecht, the Netherlands}
\altaffiltext{11}{Istituto Fisica Cosmica e Applicazioni all'Informatica (CNR), Via Ugo La Malfa 153, 90146 Palermo, Italy}
\altaffiltext{12}{Istituto Fisica Cosmica e Tecnologie Relative (CNR), Via Bassini 15, 20133 Milan, Italy}
\altaffiltext{13}{Columbia Astrophysics Laboratory, Columbia University, New York, NY 10027, U.S.A.}

\begin{abstract}
GRB~991014 is one of the shortest gamma-ray bursts detected so far with the 
Wide Field Cameras aboard BeppoSAX, both in $\gamma$-rays and X-rays. The
duration is 9.6~s in 2-28 keV and 3.2~s in 40 to 700 keV (as measured 
between the times when 5 and 95\% of the burst photons have been accumulated).
We refine the
{\em InterPlanetary Network\/} annulus of the burst, present the 
detection of the X-ray afterglow of GRB~991014 within this refined annulus,
and discuss X-ray and $\gamma$-ray
observations of the prompt and afterglow emission. Except for the briefness
of the prompt event, no other unusual aspects were found in the prompt and
afterglow observations as compared to such measurements in previous
gamma-ray bursts.
\end{abstract}


%==============================================================================

\section{Introduction}
\label{secintro}

The Wide Field Cameras (WFCs) on board the BeppoSAX satellite 
have, since launch in April 1996, detected and localized over thirty gamma-ray
bursts (GRBs). The accurate (within a few arcminutes) and swift
(within a few hours) localizations have enabled the first-time detections
of multi-wavelength afterglow emission. The optical spectra of eight
such afterglows revealed red-shifted imprints of identifiable spectral
features with $z$'s ranging
between 0.43 (GRB~990712, Galama et al. 1999) and 3.42 (GRB~971214,
Kulkarni et al. 1998), with one exceptional redshift of $z=0.0085$
if GRB~980425 is associated with SN1998bw (Galama et al. 1998).

GRB~991014 triggered the Gamma-Ray Burst
Monitor on BeppoSAX at 1999, Oct.~14.911508 UT (Tassone et al. 1999).
It was simultaneously detected with WFC unit 2 at an off-axis angle of
15$^{\rm o}$ in a total field of view of 40$^{\rm o}\times40^{\rm o}$.
The detection was of low significance and the satellite attitude
configuration was poorly determined so that the error radius of
the GRB location was relatively large at 6\arcmin. The location
was communicated about 9~hrs after the burst. Within one day after
the burst, a preliminary {\em InterPlanetary Network\/} (IPN) annulus
was published (Hurley \& Feroci 1999) which intersects the WFC error
region and reduces the size of the error region by 65\%. Optical and
radio follow-up
observations were performed but no afterglow was detected at
these wavelengths. The sensitivity limit in the optical $R$-band is $23.1$~mag
for observations starting 11.2~hrs after the burst (Thorstensen et al.
1999; Uglesich, Halpern \& Thorstensen 1999) and at radio wavelengths
125~$\mu$Jy (8.46~GHz) and 220~$\mu$Jy (1.4~GHz)
for observations starting 1.68~d after the burst (Taylor, Frail \& Kulkarni
1999).

GRB~991014 also triggered the Burst And Transient Source Experiment (BATSE)
on the Compton Gamma-Ray Observatory (trigger no. 7803, Giblin, Kippen
\& Sahi 1999). As measured in 50-300 keV peak flux at 64~ms resolution,
the burst is in the top 32\% of the BATSE bursts. In fluence ($>$25~keV), it
ranks in the top 58\%. The T90 duration (i.e., the time interval that
encompasses 90\% of the detected photons above the background starting from
when 5\% is accumulated) according to these data is 
$4.67\pm0.47$~s. The hardness ratio, as determined from the ratio of the
fluence in BATSE channels 3 (100-300~keV nominal, fluence in erg~cm$^{-2}$)
to that in channel 2 (50-100~keV),  is $H_{32}=1.36\pm0.13$.

GRB~991014 is, within the set of WFC-detected bursts, the shortest in
$\gamma$-rays which was followed up with sensitive X-ray observations
(GRB~980326 has approximately the same duration as GRB~991014 but
was not followed up). Nevertheless, it is unlikely that GRB~991014 is
a member of the class of short GRBs that have been recognized for many years 
(Norris 1984, Hurley 1992),
that tend to have harder energy spectra than the
long ones (Dezalay et al. 1992, 1996; Kouveliotou et al. 1993), 
and from which so far no afterglow has been detected in any wavelength
regime. Giblin et al. (1999) identify GRB~991014 as belonging to the
class of long bursts based on the BATSE-determined
duration and spectral hardness.

We present $\gamma$-ray and X-ray measurements, both of the burst event
and the afterglow of GRB~991014.

%==============================================================================
\section{Instrumentation and X-ray follow-up observations}
\label{sectionobs}

Our GRB and X-ray afterglow measurements were carried out with 
the following instruments on board BeppoSAX (Boella et al. 1997b).
The Gamma-Ray Burst Monitor (GRBM; Amati et al. 1997 and Feroci et al. 1997) 
comprises 4 lateral shields of the Phoswich
Detector System (PDS, Frontera et al. 1997)
and has a bandpass of 40 to 700 keV. The normal 
directions of two shields are each co-aligned with the viewing direction of a 
WFC unit. The GRBM
has 4 basic data products per shield for a GRB: a time history of the 40 to 
700 keV intensity with a variable time resolution of up to 0.48~ms, 
1~s time histories in 40 to 700 and $>$100 keV, and a 256-channel
spectrum accumulated each 128~s (independently phased from GRB
trigger times; 240 of these channels contain scientific data up to 650~keV).
The WFC instrument (Jager et al. 1997) consists of two identical
coded aperture cameras 
each with a field of view of 40$^{\rm o}\times40^{\rm o}$ full-width to zero 
response and an angular resolution of about 5\arcmin. The bandpass is 2 to 28 
keV. 
The narrow field instruments (NFI) consist of the low-energy -- 0.1 to 10 keV --
and the medium-energy -- 2 to 10 keV-- concentrator spectrometer (LECS and MECS
respectively, see Parmar et al. 1997 and Boella et al. 1997a respectively),
the PDS (15 to 300 keV), and the high-pressure gas scintillation
proportional counter (4 to 120 keV, Manzo et al. 1997).

GRB~991014 was declared a target of opportunity (TOO) for the 
NFI and a follow-up observation was performed between October 15.4
(13~h after the burst) and October 16.4 UT (we will refer to this 
observation as TOO1). The exposure time for the MECS was 38.1~ks. 
Two sources were identified in MECS data that are within the combined
WFC/IPN error region (In~'t~Zand et al. 1999a). Neither of them showed
clear variability. Therefore, a second observation was carried out
(TOO2) between October 24.8 and 25.9 UT with a MECS exposure time
of 48.2~ks.
The X-ray afterglow was identified with the one source that had
disappeared in 
this second observation: SAX J0651.0+1136 (In~'t~Zand et al. 1999b). The
burst is at a fairly low Galactic latitude with
$(l^{II},b^{II})$=(202\fdg5,+5\fdg2).

%==============================================================================
\section{Analysis}
\label{sectionana}

\subsection{The burst event}
\label{subsectionburst}

\begin{figure}[t]
  \begin{center}
\epsscale{0.8}
\plotone{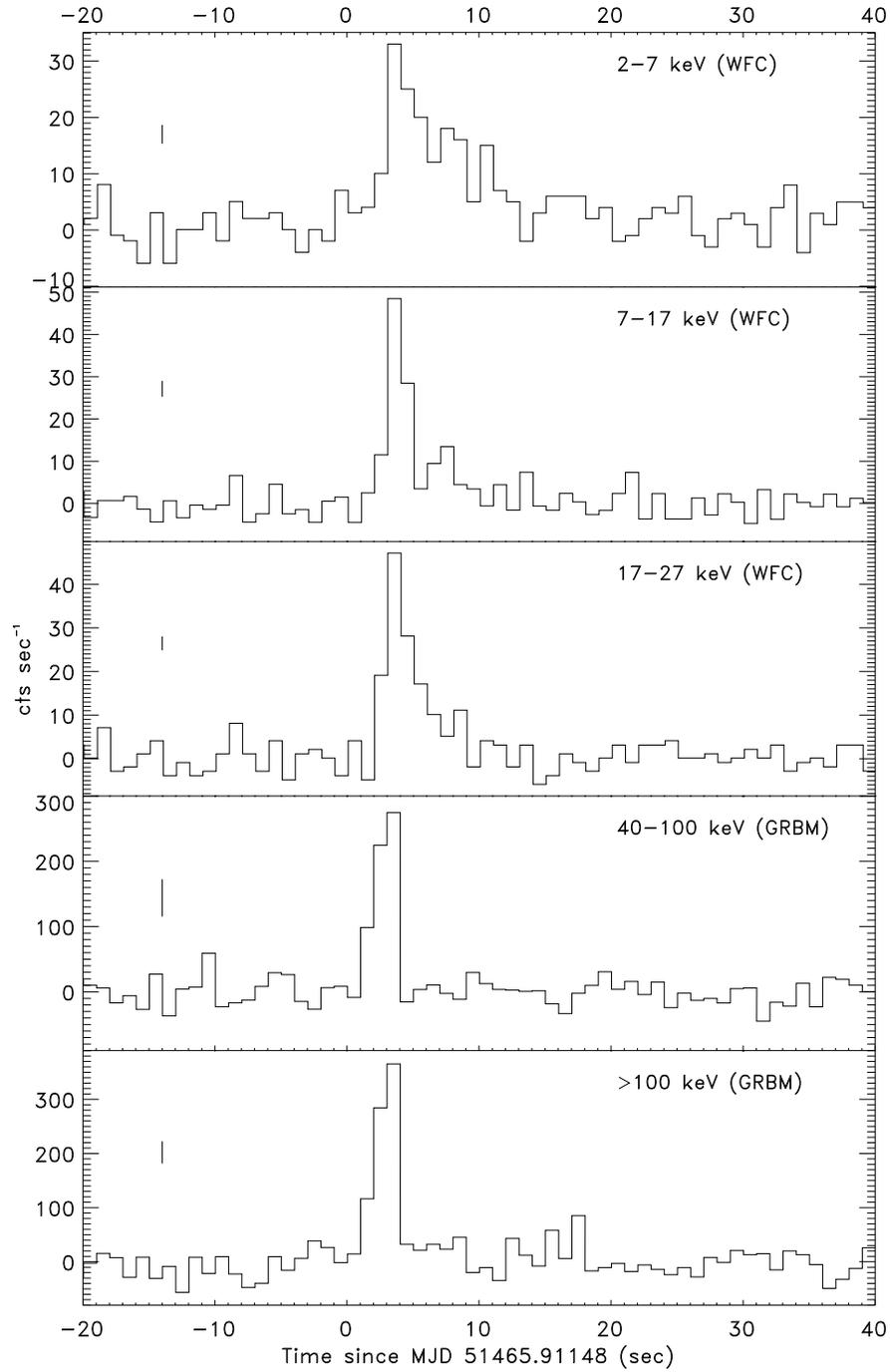}
  \caption{Time history of the burst itself as seen with WFC and GRBM, 
at a time resolution of 1~s. The vertical bars near the left edge
of each panel indicate the 1$\sigma$ errors in the rates.\label{figgrblc}
}
  \end{center}
\end{figure}

In figure~\ref{figgrblc}, the time profile of the burst is shown in various
bandpasses. The profile is comparatively simple with a single peak at
all energies. Within 1~s, the peak occurs simultaneous at all energies.
The peak profile appears to flip from below to above 40~keV
(i.e., it has a fast-rise exponential-decay profile below 40~keV and
an exponential-rise fast-decay one above that). The burst lasts longest at
the lowest energies. In the 2-28 keV band the T90 duration is $9.6\pm1.3$~s,
while in 40 to 700 keV this is $3.2\pm0.7$~s. We generated a spectrum
from the 7-s interval when the WFC signal was brightest. 
This spectrum can be modeled by a power law function 
$N(E)(:)E^\Gamma$ phot~s$^{-1}$cm$^{-2}$keV$^{-1}$
with photon index $\Gamma=-0.6\pm0.2$ assuming an interstellar
absorption of $N_{\rm H}=2.5\times10^{21}$~cm$^{-2}$ (according to
an interpolation of the HI maps by Dickey \& Lockman 1990). The peak
flux in 1-s time resolution following this spectrum is 
$(3.2\pm0.3)\times10^{-8}$~erg~s$^{-1}$cm$^{-2}$ (2-10 keV). We 
estimate the 2-10 keV fluence at $1.0\times10^{-7}$~erg~cm$^{-2}$.
For 2 to 28 keV these numbers should be multiplied by 4.1.

The WFC observation on this field was ongoing for 1.15~d before the
occurrence of the GRB. A total of 39.5~ksec net exposure time was accumulated.
The GRB was not detected during this time, with a 3$\sigma$ flux upper limit
of $1.2\times10^{-10}$~erg~s$^{-1}$cm$^{-2}$ (2-10 keV, assuming the same
spectral shape). Beyond 8~s after the burst onset, only 972~s of net
exposure time were left in the observation. We could not find any signal
from the burst position on time scales of 5~s, 10~s, or 972~s, and
in 2-10 or 2-28 keV. The 3$\sigma$ upper limit for 972~s in 2-10 keV is 
$7\times10^{-10}$~erg~s$^{-1}$cm$^{-2}$.

In 40 to 700 keV, the peak
intensity is $(6.4\pm0.5)\times10^2$~c~s$^{-1}$ in a time resolution of
1~s and twice as large, $(1.2\pm0.2)\times10^3$~c~s$^{-1}$, in a resolution
of 62.4~ms. The 1~s peak flux is
$(4.2\pm0.5)\times10^{-7}$~erg~s$^{-1}$cm$^{-2}$
and the fluence $(9\pm1)\times10^{-7}$~erg~cm$^{-2}$. This brings the
X-ray to $\gamma$-ray fluence ratio to 11\% which is relatively high though
not exceptional.
A Fourier power spectrum of the high time resolution data of
the GRBM shows no significant features between $3$~Hz and 64~Hz.

Beyond 40 keV the burst shows no spectral evolution, although the statistical
quality does not permit a sensitive analysis. 
The photon index as derived from a power law fit to the two-channel 1~s GRBM
data shows no evolution with $\Gamma$ ranging from $-2.0\pm0.7$ to $-1.9\pm0.3$
over the three brightest time bins, but the longer duration in the WFC band
with respect to the GRBM band suggest an overall hard-to-soft evolution. 
The difference in $\Gamma$ between the GRBM and WFC data
suggests a break in the spectrum between the WFC and GRBM bands.

The statistical quality of the 256-channel GRBM data of GRB~991014 does not
permit a sensitive spectral analysis.

\subsection{A refined IPN annulus}

The receipt of final Ulysses ephemeris and timing data, and of the final
BeppoSAX ephemeris data, has allowed us to reduce the width of the
Ulysses/GRBM IPN annulus from 3\farcm3 (Hurley \& Feroci 1999) to
1\farcm5 (3 sigma).  The final annulus, fully contained within the old
one,
is shown in figure~\ref{fignfimap}.  The initial error box, formed
by the intersection of the preliminary IPN annulus with the WFC error circle,
had an area of 36 square arcminutes.  The final error box,
formed by the intersection of the refined IPN annulus with the
final NFI error circle, has an area of approximately 4 square
arcminutes. 

\subsection{The X-ray afterglow}

The X-ray field around GRB~991014 is somewhat complicated. Several
faint X-ray sources can be recognized in a circular field of diameter 
25\arcmin\ around the WFC position. Because of this complication,
we carried out a maximum-likelihood analysis of the MECS
imaging data. In comparison to the standard analysis methods, the
maximum-likelihood method is able to fully take into account the point
spread function (PSF) which 1) is critical in disentangling sources that are
so close to each other that their PSFs overlap considerably, 2) ensures
that all photons detected from each point source are retrieved, and
3) takes into account the Poissonian nature of the data. The maximum-likelihood
analysis was performed in two steps. In the first step a map was generated
of detection significance, and in the second step the flux and position of
all point sources were extracted. For details we refer to the Appendix.
The significance map for TOO1 data is given in figure~\ref{fignfimap}.

For the combined MECS TOO1 and TOO2 imaging data, four
point sources are required to satisfactorily describe the data. Their
positions are indicated in figure~\ref{fignfimap}.
The elongated structure within the initially reported as well as refined
IPN/WFC error box is explained by two point sources separated by 1\farcm5.
We designate the upper (most northern) source as S1 and the lower one as S2.

The photon count numbers were determined in various time intervals
by maximum-likelihood fits of the 4 point source strengths -- while keeping
the positions fixed -- and the background level.
The average MECS (units 2 and 3 added) photon count rate during TOO1 is 
$(3.5\pm0.4)\times10^{-3}$~c~s$^{-1}$ for S1 and 
$(2.2\pm0.4)\times10^{-3}$~c~s$^{-1}$ for S2. We divided TOO1 in 3
time intervals to search for variability of the two sources, without a
definite result (the maximum difference in photon count rate is for S1,
but only at 1.4$\sigma$). In TOO2 the average rates are
$<1.2\times10^{-3}$~c~s$^{-1}$ for S1 (3$\sigma$ upper limit) and
$(2.5\pm0.3)\times10^{-3}$~c~s$^{-1}$ for S2. From these latter numbers we
identify S1 as the afterglow source of GRB~991014. S1 is equal to
SAX~J0651.0+1136. Its position is
$\alpha_{2000}$=6$^{\rm h}$51$^{\rm m}$02.9$^{\rm s}$, 
$\delta_{2000}$=+11$^{\rm o}$36\arcmin03\arcsec. The statistical error
radius is 40\arcsec\ (99\% confidence) but the systematic error enlarges
the final error radius to 1\farcm5. The combined IPN/NFI error box is
depicted in figure~\ref{fignfimap}.
The ratio of the upper
limit in TOO2 and the average flux in TOO1 implies that the
decay index is steeper than $-0.4$.

A spectrum was generated for the afterglow from the LECS and MECS TOO1 data by
maximum-likelihood fits of the 4 point source strengths and background
level in four logarithmically-sized bands between 0.2 and 2.2 keV for
the LECS data and eight logarithmically-sized bands between 1.6 and 10~keV
for the MECS data (2 units).
A simple absorbed power law model with Galactic interstellar absorption of
$N_{\rm H}=2.5\times10^{21}$~cm$^{-2}$ (following Dickey \& Lockman 1990)
fits the data well ($\chi^2=5.94$ for 6 dof) with a photon index of
$\Gamma=-1.53\pm0.25$.  The average (unabsorbed) flux is 
$(3.5\pm0.5)\times10^{-13}$~erg~s$^{-1}$cm$^{-2}$ (2-10 keV) which is
a factor of 10$^5$ smaller than the 2-10 keV peak flux of the gamma-ray burst
itself. This implies that the energy emitted in the X-ray afterglow is
comparable to that in the burst itself, and a decay index between the WFC
and the MECS fluxes of $-1.0\pm0.2$ which is not extraordinary quick when
compared to other cases (e.g., Piro 2000). The data provide a constraint
on $N_{\rm H}$: it is less than $8.6\times10^{21}$~cm$^{-2}$.

No signal was seen in the PDS data of TOO1. The 2$\sigma$ upper limit in
15-100 keV is 30 times the flux predicted by an extrapolation of the 
MECS-measured spectrum.

The NFI field seems complicated, with a concentration of sources near the
GRB position. Actually this number of sources is not a surprise given
the number of background sources expected. Giommi, Fiore \& Perri (1999) have
estimated log$N(>S)$-log$S$ distributions for 2-10 keV background sources
in BeppoSAX observations at high Galactic latitudes and find that
for a flux limit of $S=1.0\times10^{-13}$~erg~s$^{-1}$cm$^{-2}$ (2-10 keV)
there are approximately 10 sources per square degree. The chance probability
to find 3 or more sources in a 25\arcmin-diameter circle is 18\%. The Galactic
latitude is fairly low for GRB~991014. Therefore, one may actually expect
more sources than those already counted at high Galactic latitude,
particularly above 2 keV given $N_{\rm H}=2.5\times10^{21}$~cm$^{-2}$.
The concentration towards the GRB may be a selection effect because the
sensitivity of the MECS increases towards the center position due to
vignetting of the concentrators.

%==============================================================================
\section{Conclusion}
\label{sectiondis}

The constraints that could be derived for the indices of the spectrum and 
temporal decay are not unusual as compared to the X-ray afterglows of other
GRBs (for a review of X-ray afterglows, see for example Piro 2000). Also,
the spectrum of the prompt emission is not unusual.
The upper limit on the optical afterglow is fairly faint. This is not
the result of Galactic extinction which in this direction is a mild
$A_V\simeq1.4$, based on HI maps by Dickey \& Lockman (1990). It may
be the result of extinction in a host galaxy. We remark that it is not
exceptional that no optical afterglow was detected. This applies to
about half of all well-localized GRBs (e.g., Kulkarni et al. 2000).

We conclude that GRB~991014 does not exhibit behavior in the prompt
spectrum or the afterglow that would make it distinct from other {\em longer}
bursts.

%------------------------------------------------------------------------------

\acknowledgements
We are grateful to the staff of the BeppoSAX Scientific Operation Center,
the Mission Planning Team and the Science Data Center for their support in
obtaining and processing the data. We thank Tim Giblin for useful 
discussions. KH is grateful for Ulysses support
under JPL Contract 958056, and to the BeppoSAX guest investigator
program and NASA grant NAG5-9126 for IPN support. The BeppoSAX satellite
is a joint Italian and Dutch program.

%==============================================================================

%==============================================================================
\appendix
\section*{APPENDIX}

The image analysis is based on maximum-likelihood ratio (MLR)
tests performed at certain user-defined grid points in a sky field around the
object of interest, to search for point sources on top of
a flat background. At each grid point ($x_{sky},y_{sky}$) one determines
the maximum likelihood under two hypotheses: ${\cal{H}}_0$ is a description
of the data in terms of the flat background model only and ${\cal{H}}_1$
a description in terms of the flat background model and a point source at 
($x_{sky},y_{sky}$). Under ${\cal{H}}_1$, the number of counts 
$\mu_{ij}$ expected in a measured sky pixel $(i,j)$ is given by :

\begin{equation}
\mu_{ij}^{{\cal{H}}_1} = b + s\cdot f_{ij} 
\end{equation}

\noindent where $b$ and $s$ represent the (flat) background
level and source strength respectively, and $f_{ij}$ the number of
source counts in pixel $(i,j)$ from a source at ($x_{sky},y_{sky}$) (see
Boella et al. 1997a). The measured number of counts in pixel $(i,j)$ is 
$N_{ij}$.

By minimizing the natural logarithm of the likelihood ${\cal{L}}$ under
${\cal{H}}_1$, given by $\ln({\cal{L}}_{{\cal{H}}_1}) = \sum_{i}\sum_{j} 
\left( N_{ij} \ln(\mu_{ij}^{{\cal{H}}_1}) - \mu_{ij}^{{\cal{H}}_1}\right)$,
with respect to its free scale parameters, $b$ and $s$, one can
derive the flux and flux uncertainty from $s$ and its error for a
putative source at position ($x_{sky},y_{sky}$). From optimizations under
${\cal{H}}_1$ and ${\cal{H}}_0$ one can determine the likelihood ratio
$\lambda$ defined as $-2\ln({\cal{L}}_{{\cal{H}}_0}/{\cal{L}}_{{\cal{H}}_1})$. 
This quantity is distributed as $\chi^2$ for 1 degree of freedom for a known
source position (e.g., Eadie et al. 1971) and yields the source detection
significance. The
maximum-likelihood ratios evaluated at the grid points ($x_{sky},y_{sky}$)
constitute the MLR map.

\begin{figure}[t]
  \begin{center}
\plotone{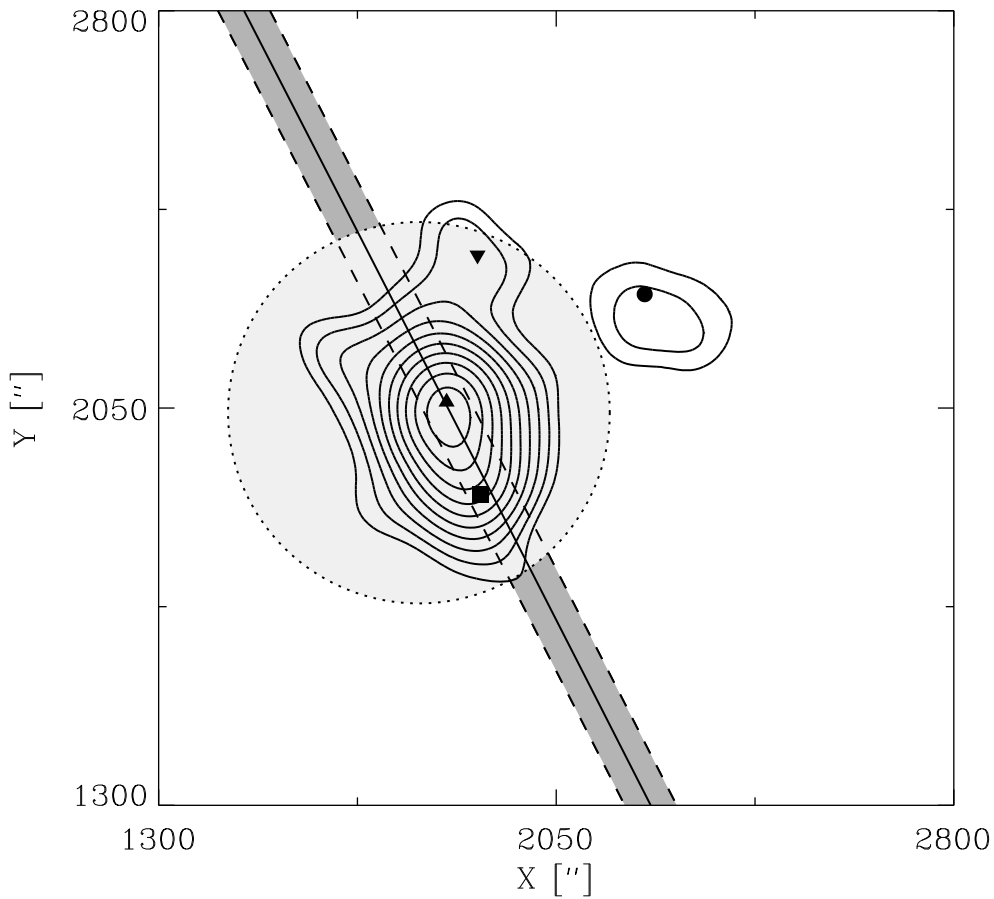}
  \caption{Maximum-likelihood contour map of the MECS data from TOO1. 
The contour levels start at the equivalent of a 3$\sigma$ detection
and increment with steps of 1$\sigma$ (1 d.o.f.). The highest contour has a
level of 12$\sigma$. The X and Y coordinates are with respect to the
instrument coordinate frame. The symbols refer to the fitted positions of
the four significant point sources detected in the combined
TOO1 and TOO2 data. The upward
pointed triangle indicates the afterglow of GRB~991014 (SAX~J0651.0+1136).
The light-shaded
circle is the WFC error region of GRB~991014 (99\% confidence) 
and the dark-shaded area
between the two
parallel lines bound the IPN annulus
(3$\sigma$ confidence)\label{fignfimap}
}
  \end{center}
\end{figure}

The MLR map for the MECS TOO1 imaging data (8\arcsec\ grid point spacing; no
energy selection applied) is given in figure~\ref{fignfimap}. The maximum value
corresponds to a $\sim 12\sigma$ detection significance, and is located
within the cross section of the WFC and IPN error regions.
From the map it is clear that multiple sources contribute to the observed
excess. For instance, the $\sim 12\sigma$ excess is elongated and suggestive
of two or more close point sources. In order to determine the minimum number
of point sources on a flat background that are required to satisfactorily
describe the measured data, we applied
a maximum-likelihood source-fitting procedure. In this procedure the data
are described in terms of a flat background model and a number of sources
each with a {\em free} position. One starts with a model composed of the flat
background model and a single source at an arbitrary {\em free} position
and optimizes the likelihood with respect to the background level, source 
strength and source-position parameters simultaneously (4 free 
parameters). Next one determines the likelihood improvement with respect to
the zero hypothesis ${\cal{H}}_0$ that the data are described 
in terms of a flat background model only (1 free parameter).
If the improvement exceeds the $3\sigma$ threshold for 3 $(= 4 - 1)$ d.o.f.
one continues and compares the optimized likelihood of a model composed of a
flat background and 2 sources at arbitrary positions with the optimized 
likelihood for the previous model with 1 source at an arbitrary position.
This process is repeated until the likelihood improvement is below the
$3\sigma$ threshold for 3 d.o.f. At each iteration one obtains the optimum 
background scaling parameter and for each source its strength and position
together with uncertainty estimates on each free parameter.

%==============================================================================

\end{document}